\documentstyle[12pt]{article}
\def\dfrac{\displaystyle\frac}
\def\i{\imath}
\def\g{\gamma}
\def\de{\delta}
\def\a{\alpha}
\def\l{\lambda}

\def\hp{\hat{p}}
\def\hk{\hat{k}}

\def\bp{{\bf p}}
\def\bx{{\bf x}}

\def\bq{{\bf q}}
\def\d{\partial}

\def\noi{\noindent}
\def\ve{\varepsilon}

\newcommand{\Eq}[1]{Eq.~(\ref{#1})}

\newcommand{\refc}[1]{Ref.~\cite{#1}}
\newcommand{\refs}[1]{Refs.~\cite{#1}}
\newcommand{\bea}{\begin{eqnarray}}
\newcommand{\eea}{\end{eqnarray}}
\newcommand{\be}{\begin{equation}}
\newcommand{\ee}{\end{equation}}
\newcommand{\bc}{\begin{center}}
\newcommand{\ec}{\end{center}}
\newcommand{\ba}{\begin{array}}
\newcommand{\ea}{\end{array}}

\newcommand{\cL}{{\cal L}}

\newcommand{\annp}[3]{{\it  Ann. Phys. (N.Y.) }{{\bf #1} {(#2)} {#3}}}
\newcommand{\ijmp}[3]{{\it Int. J. Mod. Phys. } {{\bf #1} {(#2)} {#3}}}
\newcommand{\jp}[3]{{\it J. Phys. } {{\bf #1} {(#2)} {#3}}}
\newcommand{\fp}[3]{{\it Fortschr. Phys. } {{\bf #1} {(#2)} {#3}}}

\newcommand{\np}[3]{{\it  Nucl. Phys. }{{\bf #1} {(#2)} {#3}}}

\newcommand{\prd}[3]{{\it  Phys. Rev. D} {{\bf #1} {(#2)} {#3}}}
\newcommand{\prl}[3]{ {\it Phys. Rev. Lett.}{{ \bf #1} {(#2)} {#3}}}
\newcommand{\pl}[3]{{\it  Phys. Lett. }{{\bf #1} {(#2)} {#3}}}

\newcommand{\sovjnp}[3]{{\it Sov. J. Nucl. Phys. }{{\bf #1} {(#2)} {#3}}}

\normalsize
\begin{document}

\thispagestyle{empty}
\phantom{q}
{}
\vspace{0.5cm}
\thispagestyle{empty}
\begin{flushright}                              FIAN/TD/97-16\\
                                                hep-th/9711137\\
                                                October 1997

\end{flushright}
\vspace{3.5cm}

\bc
\normalsize


{\LARGE\bf Gauge Invariance of QED$_{2+1}$}

\vspace{4ex}

{\Large I.~V.~Tyutin$^{\dagger}$ and Vad.~Yu.~Zeitlin$^{\ddagger}$}

\vspace{2ex}
{\large I.~E.~Tamm Theory Department,  P.~N.~Lebedev Physical Institute

\vspace{1ex}
Russia, 117924, Moscow, Leninsky prospect, 53}
\vspace{5ex}

\ec

\normalsize
\begin{quote}
The problem of gauge invariance of the physical sector of
(2+1)-dimensional Maxwell-Chern-Simons quantum electrodynamics (QED$_{2+1}$)
is studied. It is shown that using Proca mass term for the infrared
regularization one obtains gauge-invariant fermion mass and the physical mass
shell of QED$_{2+1}$ is well-defined in all orders of the perturbation
theory.
We are demonstrating also a class of gauges in the framework of QED$_{2+1}$,
including transversal and Feynman-like ones, where the physical sector is
well defined and independent of the gauge parameter.
\end{quote}

\vfill
\noindent
$^\dagger$ E-mail address: tyutin@lpi.ac.ru

\noindent
$^\ddagger$ E-mail address: zeitlin@lpi.ac.ru

\newpage


\setcounter{page}{1}

\section{Introduction}

Recent years have witnessed a great interest to field theory models in the
space-time of dimension (2+1) (see e.g. Refs. [1--7] and references to [1]).
These models arise as a temperature reduction of
four-dimensional models \cite{AP,KKC} or may be used for describing
low-dimensional models like films and wires
\cite{W,LSW91,QHE}. Besides, field theories in low-dimensional space-time may
be of considerable interest as being simplified (toy) models of more realistic
(and more complicated) four-dimensional models.

Many unusual properties of 3-dimensional models are due to the
gauge-invariant mass of the gauge field, which is not induced by spontaneous
symmetry breaking. The mass may be introduced by adding
the Chern-Simons term to the bare action [1, 12--14]
(topologically massive gauge theories) and/or be generated dynamically via
interaction with fermions \cite{DJT}.

Low-dimensional field theories have severe infrared singularities. In
\refs{DJT,JT} where (2+1)-dimensional quantum electrodynamics
(QED$_{2+1}$) was first studied it was shown that the singularities lead
to dependence of the one-loop physical mass of the fermion on the gauge
parameter in an arbitrary $\a$-gauge, and the higher order
perturbative diagrams are apparently divergent. It was assumed in the
above papers that one should use transversal (Landau) gauge in the perturbative
calculations, since the photon propagator is less infrared singular in this
case and infrared divergencies  may be absent from the Green functions.

Nevertheless, it is not clear whether gauge variance of the physical mass
of the fermion indicates an anomaly (of Adler's anomaly type) even in
the transversal gauge. The latter problem may be raised in other way: whether
the theory is unitary in the subspace of states of fermions and massive
photons?  As far as we know \cite{DJT,DFGGS}, the physical unitarity of
QED$_{2+1}$ in the transversal gauge has been stated by comparing it to the
Coulomb gauge where the nonphysical sector is absent. The equivalence of the
two gauges has been shown by modifying longitudinal terms in the gauge field
propagator. This does not affect the Green function on the physical mass
shell, but transforms the Coulomb photon propagator into that in the
transversal gauge.  Unfortunately, the photon propagator is ill-defined
in the Coulomb gauge because of $1/{\bf p^2}$ infrared singularities, thus the
latter gauge requires an additional definition.

In this paper we present a detailed analysis of unitarity, both by introducing
appropriate infrared regularization -- by  adding the Proca mass term to the
bare action (this modification of QED$_{2+1}$ we shall call the Proca model)
and in the framework of QED$_{2+1}$ in a class of gauges including
transversal and Feynman-like ones. In the Proca model all the photon modes
are massive and infrared divergencies are absent (however, just two modes are
 physical).  We shall show that the observables are independent of the gauge
 parameter and after the infrared regularization is removed (by
letting the Proca mass tend to zero) the scattering amplitudes involving one
of the photons (that with vanishing mass) die off, and the resulting theory
is just QED$_{2+1}$ in the physical subspace of the transversal gauge.
Furthermore, the unitarity in the physical sector is demonstrated directly in
the framework of QED$_{2+1}$.

The paper is organized as follows. In Sec.~2 we are showing the reasons why
 the fermion mass in QED$_{2+1}$ is gauge dependent despite the Ward
identities are explicitly satisfied. In Sec.~3 we are demonstrating by
analyzing the corresponding Ward identities that addition of the Proca mass
term to QED$_{2+1}$ Lagrangian removes dependence of the physical sector on
the gauge parameter. Gauge-variant mode of the electromagnetic field splits
off and does not contribute to any scattering amplitude, while vertex
functions are gauge-invariant on the fermion mass shell. The limit of zero
Proca mass is studied and it is shown that the regularization may be taken off.
In Sec.~4 the unitarity and gauge independence of the physical sector
in a class of gauges is shown in the framework of QED$_{2+1}$.
In the Appendix the normalization conditions in the theory with the
Chern-Simons term are obtained.

\section{Gauge variance of QED$_{2+1}$}

First, let us consider the reasons for gauge dependence of the fermion
mass  (generally speaking, any observable) in QED$_{2+1}$.

The Lagrangian of QED$_{2+1}$ is the following:

        \be
        \cL_{(0,\a)} = -\frac14 F_{\mu\nu}F^{\mu\nu} +
        \frac{\theta}4 \ve_{\mu\nu\alpha}F^{\mu\nu}A^\alpha
        - \frac1{2\alpha} (\d_\mu A^\mu)^2+
        \bar{\psi}
        (\imath \hat{\partial} + e \hat{A} -M)\psi~~~.
        \label{mcs}
        \ee

\noi
The first term in the left-hand side of \Eq{mcs} is the Maxwell term
($F_{\mu\nu}=\d_\mu A_\nu - \d_\nu A_\mu$), the second one is
so-called Chern-Simons term, the third one is the gauge fixing term,
and the last one describes fermions and interaction.  We shall call the
above model QED$_{2+1}$ in $\alpha$-gauge.

In a standard way one may obtain the Ward identities for
generating functional of vertex functions $\Gamma$:

        \be
        \d_\mu \dfrac{\delta\bar{\Gamma}}{\delta A_\mu}=
        \i e \psi \dfrac{\delta\bar{\Gamma}}{\delta \psi}
        -\i e \bar{\psi} \dfrac{\delta\bar{\Gamma}}{\delta \bar{\psi}}~~~,
        \label{ward}
        \ee

\noi
where

        \be
        e^{\i W(I,\eta,\bar{\eta})}=\int D\!A D\!\psi D\!\bar{\psi}
        e^{\i \int d\!x ({\cal L} +IA +\eta\psi + \bar{\eta}\bar{\psi})}~~~,
        \ee

        \be
        \Gamma = W - IA - \eta\psi - \bar{\eta}\bar{\psi}
        \equiv \bar{\Gamma} - \frac1{2\alpha} (\d A)^2~~~,
        \ee

        \be
        A_\mu = \frac{\de W}{\de I^\mu}~, \quad
        \psi = \frac{\de W}{\de \eta}~,\quad
        \bar{\psi} = \frac{\de W}{\de \bar{\eta}}\quad.
        \nonumber
        \ee

Taking into account \Eq{ward} one has the following expression for the
variation of the $\bar{\Gamma}$ with respect to the gauge parameter:

        \be
        \frac{\delta \bar{\Gamma}}{\delta \alpha} =
        \frac{e}{2\alpha}
        \int d\!x d\!y
        \left[
        \d_\mu \frac{\de^2W}{\de I^\mu(x)\de {\eta}(y)}
        {\rm D}_{(0)} (x-y) \frac{\delta \bar{\Gamma}}{\delta {\psi}(y)}
        -
        \d_\mu \frac{\de^2W}{\de I^\mu(x) \de \bar{\eta}(y)}
        {\rm D}_{(0)} (x-y) \frac{\delta \bar{\Gamma}}{\delta \bar{\psi}(y)}
                                                \right]~~,
        \label{dgda}
        \ee

        \be
        {\rm D}_{(m^2)} (x-y) = \int \frac{d\! p}{(2\pi)^3}
        \frac{e^{-\i p(x-y)}}{p^2-m^2}~~~.
        \label{rmd}
        \ee

If only vertex-function-type diagrams (i.e. one-particle-irreducible
diagrams) were nonsingular when the external momenta are on the mass
shell, then by using the standard arguments employed to prove the theorem of
the equivalency and gauge invariance \refc{KT}, one could show that
the transversality of the vertex functions (except
two-point function, the inverse photon propagator) in photon momenta
with fermion momenta on the mass shell follows from  \Eq{ward}, and
independence of the vertex functions (except two-point photon vertex) on the
gauge parameter $\a$ appears from \Eq{dgda}.

Unfortunately, in QED$_{2+1}$ in $\alpha$-gauge due to the ill infrared
behavior of the longitudinal part of the gauge field propagator
$D^{(0,\alpha)}_{\mu\nu}(p)$,

       \be
        D^{(0,\alpha)}_{\mu\nu}(p) =\frac{-\i}{p^2 - \theta^2 +\i\ve}
        \left(
                \left( g_{\mu\nu} - \frac{p_\mu p_\nu}{p^2} \right)
                + \i \theta \ve_{\mu\nu\alpha}\frac{p^\alpha}{p^2}  \right)
                        -\i\alpha \frac{p_\mu p_\nu}{p^4}~~~,
        \label{dmunu_mcs}
        \ee

\noi
one-particle irreducible diagrams may be singular on the mass shell, thus
the gauge invariance cannot be proven despite validity of Eqs.~(\ref{ward})
and (\ref{dgda}). Consider fermion mass operator as an illustration.
It follows from \Eq{dgda} that

        \bea
        \lefteqn{\frac{\d \Sigma (x-y)}{\d \a} =
        -\i \frac{e}2 \int d\!z
        \left[
        \int d\!z_1 d\!z_2 D_\mu (z-z_1) \Gamma^\mu(z_1,x,z_2)
        S(z_2-z)                \right]
        S^{-1} (z-y) -}\nonumber\\
        &&-\i \frac{e}2 \int d\!z
        S^{-1} (x-z)
        \left[
        \int d\!z_1 d\!z_2 D_\mu (z-z_2) S(z-z_1) \Gamma^\mu(z_1,z_2,y)
                     \right]~~~,
        \label{dsda}
        \eea

        \be
        \Gamma^\mu(x,y,z) =
        \left.
        \frac{\de}{\de A_\mu(x)}
        \frac{\de}{\de \bar{\psi(y)}}
        \frac{\de}{\de {\psi(z)}} \Gamma
        \right|_{€_\mu=\psi=\bar{\psi}=0}~~~,
         \ee

\medskip
\noi
$S$ is an exact fermion propagator (vacuum expectation value of the
$T$-product) and the mass operator $\Sigma$ is defined as follows:

        \be
        S^{-1} = S^{-1}_0 +\i \Sigma~~~,
        \ee

\noi
$S_0$ is the bare fermion propagator and the function $D_\mu$ is equal to

        \be
        D_\mu(x) = \int \frac{d\!p}{(2\pi)^3} e^{-\i px}
        \frac{p_\mu}{p^4}~~~.
        \ee

Consider \Eq{dsda} in the one-loop approximation (two terms provide the same
contribution):

        \bea
        \lefteqn{\frac{\d \Sigma_1 (x-y)}{\d \a} =}\\
        \nonumber
        &&-\i e^2 \int d\!zS^{-1}_0(x-z) [D_\mu(x-y)S_0(x-y)\gamma^\mu]
        \equiv
        \i e^2 \int d\!zS^{-1}_0(x-z) \Sigma'(z-y)~~~.
        \eea

In the momentum space the latter is:

        \be
        \frac{\d {\Sigma}_1 (p)}{\d \a} =
        \i e^2 {S}^{-1}_0(p) \Sigma_1' (p)=
        e^2 (\hat{p}-M){\Sigma}'_1 (p)~~~,
        \label{sigma1}
        \ee

        \be
        {\Sigma}'_1 (p) = -\i \int \frac{d\!k}{(2\pi)^3}
        \frac{(\hat{p}-\hat{k}+M)\hat{k}}{k^4((p-k)^2-M^2)} =
        \frac1{8\pi}\cdot \frac1{\hp -M} + \dots~~~,
        \label{sigma'}
        \ee

\medskip
\noi
where (\dots) denotes terms which are logarithmically singular and finite on
the mass shell. The validity of Eqs. (\ref{sigma1}), (\ref{sigma'}) may
be easily checked by direct calculating of $\Sigma_1$.  However, the latter
equation does not manifest gauge invariance of $\Sigma_1$ on the mass shell
(at $\hat{p}=M$) despite the multiplier vanishing at $\hat{p}=M$, i.e. the
physical mass of the fermion is not gauge invariant. This is a consequence of
the singularity of  ${\Sigma}'_1$ on the mass shell (see
\Eq{sigma'}), which is provided by the infrared behavior of the photon
propagator in $\a$-gauge.

\section{Proca term and gauge invariance}

We shall demonstrate in this section that the gauge invariance in
QED$_{2+1}$ may be restored. As a regularization parameter we are adding the
Proca mass\footnote{(2+1)-dimensional models with the Proca term have been
considered in \refs{NS,DMcK} in a different context.} to the bare Lagrangian
\Eq{mcs}.  The Lagrangian of the Proca model is the following:

        \be
        \cL_{(m,\alpha)} = -\frac14 V_{\mu\nu}V^{\mu\nu} +
        \frac{\theta}4 \ve_{\mu\nu\alpha}V^{\mu\nu}V^\alpha + \frac{m^2}2
        V_\mu V^\mu - \frac12 (\d_\mu V^\mu)
        \frac1{\alpha} (\d_\nu V^\nu) +\bar{\psi}
        (\imath \hat{\partial} + e \hat{V}   -M)\psi ~~~,
        \label{mcss}
        \ee

        \be
        V_{\mu\nu}=\d_\mu V_\nu - \d_\nu V_\mu~~~.
        \nonumber
        \ee

\noi
We have written the gauge-fixing term in that form, since we shall use below
an arbitrary function of d'Alambertian as a gauge parameter, $\a =
\a(-\Box)$.

The solutions to the free equation of motion for the field $V_\mu$ are the
following (correct procedure of normalization of the polarization vectors is
discussed in the Appendix):

        \be
        V_\mu(x)=
        \sum_{i=1}^{3}
        \int\frac{d {\bf p}}{2\pi\sqrt{2\omega_i}}
        (e^{-\i p^{(i)} x}u^{(i)}_\mu a_{i}(\bp) + {\rm h.c.})~~~,~~~
        p_\mu^{(i)}=(\omega_i,\bp)~~~.
        \label{polar_v1}
        \ee

        \be
        u^{(1,2)}_\mu =
        \dfrac{\omega_{1,2}}
        {|\bp| \sqrt{\omega_{1,2}^2-\bp^2+m^2}}
        (p^{(1,2)}_\mu - g_{\mu 0} \frac{\omega_{1,2}^2-\bp^2}{\omega_{1,2}}
        +\i \frac{\omega_{1,2}^2-\bp^2-m^2}{\theta\omega_{1,2}}
        \varepsilon_{\mu\alpha0}p^{(1,2)\alpha})~~~,
        \label{polar_v2}
        \ee

        \be
        \omega_{1,2}=\sqrt{{\bf p}^2+m^2_{1,2}}~~~,~~~
        m_1^2= m^2+\frac{\theta^2}2+\theta\sqrt{m^2+\frac{\theta^2}4}~~~,~~~
        m^2_2=\frac{m^4}{m^2 + \frac{\theta^2}2  +
        \theta \sqrt{m^2+\frac{\theta^2}4}}~~~,
        \ee

        \be
        u^{(3)}_\mu= \frac1{m}p_\mu^{(3)} ~~~,~~~\omega_3=\sqrt{{\bf
        p}^2+m^2_3}~~~,~~~m_3^2=\alpha m^2~~~.
        \ee

\medskip
There are three excitations in the  gauge sector of the model.
Excitations with polarization vectors $u^{(1)}_\mu$ ¨ $u^{(2)}_\mu$
are massive physical modes, while the excitation with the polarization vector
$u^{(3)}_\mu$ is nonphysical with its mass dependending on the gauge parameter
(as we shall show below the latter mode is free and does not contribute to
scattering amplitudes).

The free vector field $V_\mu$ propagator is the following

	 \be
        D_{\mu\nu}^{(m,\a)}(p) =
        -\i\frac{p^2 -m^2}{(p^2 - m^2)^2-\theta^2 p^2}
        \left(
                \left( g_{\mu\nu} - \frac{p_\mu p_\nu}{p^2} \right)
                + \i \theta \ve_{\mu\nu\alpha}\dfrac{p^\alpha}{p^2-m^2}
							  \right)
               -\i\frac{p_\mu p_\nu}{p^2}
               \frac{\alpha}{p^2-\alpha m^2}~~~.
        \label{dmunu_mcss}
        \ee

\noi
It is nonsingular when $p\to 0$ for any $\alpha$. Therefore, the Proca
model is free of infrared problems. For comparison, the similar
quantities in QED$_{2+1}$ are:

        \be
        A_\mu(x)=
        \int\frac{d {\bf p}}{2\pi\sqrt{2\omega\phantom{|}}}
        (e^{-\i p^{(1)}x}e^{(1)}_\mu a_{ph}(\bp) + {\rm h.c.})
        +
        \int\frac{d {\bf p}}{2\pi\sqrt{2|{\bf p}|}}
        (e^{-\i p^{(2)}x}(e^{(2)}_\mu a_2(\bp) + e^{(3)}_\mu a_3(\bp)) + {\rm
        h.c.})~~,
        \ee

\noi
with the polarization vectors\footnote{Free equation of motion has two
linearly independent massless (nonphysical) solutions, longitudinal mode
${e}^{(2)}_\mu= c p_\mu$ and gauge-dependent mode ${e}^{(3)}_\mu=
c(|\bp| g_{\mu 0} + \frac{2\i|\bp|}{\alpha\theta}
\varepsilon_{\mu\beta0}p^\beta - \i |\bp| x_0   p_\mu)$
(in an arbitrary relativistic gauge one cannot
present all modes as plane waves, see e.g.  \refc{IZ}).
Polarizations
\Eq{polariz1} are superpositions of these modes and the coefficients are
chosen to satisfy  commutation relations
$[a_{ph}(p),a^\dagger_{ph}(q)]=[a_{2}(p),a^\dagger_{3}(q)]  =
[a_{3}(p),a^\dagger_{2}(q)]=\delta(\bp-\bq)$, other commutators are
vanishing.}:

        \be
        e^{(1)}_\mu = \frac{\omega_1}{\theta|\bp|}
        (p^{(1)}_\mu - \delta_{\mu 0} \frac{\theta^2}{\omega_1}
        +\i \frac\theta{\omega_1}\varepsilon_{\mu\alpha0}p^{(1)\alpha})
        ~~~~, ~~~ \omega_1 = p_0^{(1)}=\sqrt{\theta^2+{\bf p}^2}~~~.
        \ee

        \be
        e^{(2)}_\mu=\frac{1}{\sqrt{\theta|\bp|}} p_\mu^{(2)},~~~
        p_0^{(2)}=|{\bf p}|
        \ee

        \be
        e^{(3)}_\mu = -\frac{1}{2}\sqrt{\frac{\theta}{|\bp|}}
        \left(\alpha g_{\mu 0} + \frac{2\i}{\theta}
        \varepsilon_{\mu\beta0}p^{(2)\beta}
        + (\frac{|\bp|}{\theta^2} - \frac{\alpha}{2|\bp|}
        - \i \alpha x_0 )   p^{(2)}_\mu
                                        \right)~~~,
        \label{polariz1}
        \ee

\noi
the propagator is presented in \Eq{dmunu_mcs}.
There is only one physical excitation in the gauge-field sector of
QED$_{2+1}$, massive spin--1 photon (and two nonphysical massless modes).

One can see that the propagator
$D_{\mu\nu}^{(m,\a)}(p)$ of the Proca theory becomes
propagator of  QED$_{2+1}$ $D_{\mu\nu}^{(0,\a)}(p)$ , \Eq{dmunu_mcs} in $m
\to 0$ limit.  Then, one of the excitations of the Proca model, namely the
mode with the polarization $u^{(1)}_\mu$, becomes the physical excitation
$e^{(1)}_\mu$ of QED$_{2+1}$ in $m \to 0$ limit.

In the Proca model the Ward identities similar to \Eq{ward}
and its corollary \Eq{dgda} are:

        \be
        \d_\mu \dfrac{\delta\bar{\Gamma}}{\delta V_\mu(x)}=
        \i e \psi(x) \dfrac{\delta\bar{\Gamma}}{\delta \psi(x)}
        -\i e \bar{\psi}(x) \dfrac{\delta\bar{\Gamma}}{\delta
        \bar{\psi}(x)}~~~,
        \label{ward_alpha}
        \ee

        \bea
        \lefteqn{
        \delta \bar{\Gamma} =}
        \label{dgda_alpha}
        \\
        &&\frac{e}{2}
        \int d\!x d\!y
        \left[
        \d_\mu \frac{\de^2W}{\de I^\mu(x)\de {\eta}(y)}
        \frac{\de \a}{\a}{\rm D}_{(\a m^2)} (x-y) \frac{\delta
        \bar{\Gamma}}{\delta {\psi}(y)} -
        \d_\mu \frac{\de^2W}{\de I^\mu(x)\de \bar{\eta}(y)}
        \frac{\de \a}{\a}{\rm D}_{(\a m^2)}
        (x-y) \frac{\delta \bar{\Gamma}}{\delta \bar{\psi}(y)} \right],
        \nonumber
        \eea

        \be
        \Gamma = \bar{\Gamma} - \frac1{2} (\d_\mu V^\mu)
        \frac1{\alpha} (\d_\nu V^\nu) +\frac{m^2}2V_\mu V^\mu~~~.
        \nonumber
        \ee

Since the propagators in the Proca model are nonsingular in the infrared
limit, the arguments underlying the equivalency theorem are valid in the
theory,  thus \Eq{dgda_alpha}
implies the independence of $\bar{\Gamma}$ on $\alpha$ with the external
fermion momenta on the mass shell, and the transversality in the photon
momenta of the vertex functions with more than two external lines does follow
from \Eq{ward_alpha} (see also the next section).

One of the corollaries of the gauge invariance of the vertex function is the
fact that  the physical masses of the fermion and of
the two photon modes with polarizations $u^{(1)}_\mu$ and $u^{(2)}_\mu$ do
not depend on $\a$.  Consider one-loop corrections to the fermion mass as an
illustration.  Taking into account \Eq{dgda_alpha} one has

        \be
        \frac{\d \Sigma_1 (p)}{\d \a} = e^2 (\hp - M) {\Sigma}'_1(p)
        \label{sigma''}
        \ee

        \be
        {\Sigma}'_1 (p) = -\frac{\i}{(2\pi)^3} \int d\!k
        \frac{(\hat{p}-\hat{k}+M)\hat{k}}{(k^2-\alpha
        m^2)^2((p-k)^2-M^2)}~~~,
        \ee

        \be
        {\Sigma}'_1 (\hp=M) =  \frac1{8\pi m \sqrt{\a}}~~~.
        \ee

Since ${\Sigma}'_1(p)$ is nonsingular on the fermion mass shell, it follows
from \Eq{sigma''} that  the fermion physical mass does not depend on $\a$.
Explicit calculation of the mass operator confirms this:

        \be
        {\Sigma}_1 (p) =
        {\Sigma}^{(0)}_1 (p) + {\Sigma}^{(\alpha)}_1(p)~~~,
        \ee

        \bea
        \lefteqn{{\Sigma}^{(0)}_1 (p)=}\\
        &&-\i e^2 \int  \frac{d\!k}{(2\pi)^3}
        \frac{\g^\mu(\hat{p}-\hat{k}+M)\g^\nu}{((p-k)^2-M^2)}
        \frac{k^2 -m^2}{(k^2 - m^2)^2-\theta^2 k^2 +\i\ve}
        \left(
                \left( g_{\mu\nu} - \frac{k_\mu k_\nu}{p^2} \right)
                + \i \theta \ve_{\mu\nu\alpha}\dfrac{k^\alpha}{k^2-m^2}
                                                          \right),
        \nonumber
        \eea

        \bea
        \lefteqn{{\Sigma}^{(\alpha)}_1 (p) =
         -\i e^2 \int \frac{d\!k}{(2\pi)^3}
        \g^\mu\frac{(\hat{p}-\hat{k}+M)}{((p-k)^2-M^2)} \g^\nu
        \frac{k_\mu k_\nu}{k^2}\frac{\alpha}{k^2-\alpha m^2}=}\nonumber\\
        && -\i e^2\alpha(\hp - M) \int \frac{d\!k}{(2\pi)^3}
        \frac{(\hat{p}-\hat{k}+M)\hat{k}}{k^2((p-k)^2-M^2)
        (k^2-\alpha m^2)}~~~.
        \eea

One can see that at any $\a$ only   ${\Sigma}^{(0)}_1 $ contributes to
the mass renormalization. Therefore the mass does not depend on $\a$,
and in the $m\to 0$  limit it coincides with the mass renormalization in
QED$_{2+1}$ in the transversal gauge $\alpha=0$.

There are two important consequences of the transversality in photon momenta
of the vertex functions (with more than two external lines)
on the fermion mass shell. Since  $u^{(3)}_\mu \sim p_\mu$,
the corresponding photon mode splits off from the physical sector
which contains the fermion and two other photon modes.

Then, \Eq{polar_v2} for the polarization vector $u^{(2)}_\mu$  may be
rewritten as follows:

        \be
        u^{(2)}_\mu|_{m\rightarrow 0} =
        \frac1{m}
        p_\mu + m l_\mu(p)~~~,
        \label{limit}
        \ee

\noi
the vector $l_\mu$  has a finite limit when $m\to 0$. Therefore, due to the
transversality of vertex functions the elements of the $S$-matrix corresponding
to radiation of $n$ photons with the polarization $u^{(2)}_\mu$ (we shall call
these photons "soft") have the factor of $m^n$. In case  the vertex functions
after taking the regularization off, $m\to 0$,  have lower singularities in
$m$, $S$-matrix elements containing soft external photons vanish in this
limit, and the physical sector will contain one massive photon and the
fermion only.  Actually, it will be demonstrated in the next section that all
the vertex functions exist in QED$_{2+1}$ in the transversal gauge $\alpha=0$
on the mass shell.

Thus the analysis of the Proca model shows that after
the infrared regularization is taken off, the Feynman rules become exactly
those of QED$_{2+1}$ in the transversal gauge thus establishing unitarity of
QED$_{2+1}$ in the transversal gauge, the physical sector of the resulting
theory comprising of a fermion and a massive photon.

Note that the changing of the constant $\a$ to an arbitrary
function of d'Alambertian, $\a\rightarrow \a(-\Box)$, does not affect
the expression for the photon propagator and Ward identities
\Eq{ward_alpha} and \Eq{dgda_alpha}. Transversality of the
vertex functions in photon momenta and the fact that
$\bar{\Gamma}$ does not depend on $\a(-\Box)$ on the fermion mass shell also
hold true.  Let us choose  $\a(-\Box)$ to be

        \be
        \a(-\Box) = \xi\frac{\Box(\Box+m^2)}
        {(\Box+m^2)(\Box+(1-\xi)m^2) +\theta^2\Box}~~~.
        \ee

The corresponding photon propagator is the following

        \be
        D^{(m,\xi)}_{\mu\nu}(p)=
        -\frac{\i}{(p^2 - m^2)^2-\theta^2 p^2}
        \left(
                \left( g_{\mu\nu} - (1-\xi)\frac{p_\mu p_\nu}{p^2}
\right)(p^2-m^2) + \i \theta \ve_{\mu\nu\alpha}p^\alpha \right)~~~.
        \label{alpha0}
        \ee

In  the $m \to 0$ limit \Eq{alpha0} becomes:

        \be
        D^{(0,\xi)}_{\mu\nu}(p)=
        -\frac{\i}{p^2-\theta^2}
        \left( g_{\mu\nu} - (1-\xi)\frac{p_\mu p_\nu}{p^2}
                + \i \theta \ve_{\mu\nu\alpha}\frac{p^\alpha}{p^2}
                                                          \right)~~~.
        \label{alpha00}
        \ee

Infrared behavior of the propagator  $D^{(0,\xi)}_{\mu\nu}(p)$ is exactly
the same as that of the photon propagator in  QED$_{2+1}$ in the transversal
gauge, therefore  $D^{(0,\xi)}_{\mu\nu}(p)$ may also be used as
the QED$_{2+1}$ photon propagator. Since the physical spectrum of the Proca
model is independent of $\xi$, it will hold true for QED$_{2+1}$ as well. On
the other hand, as far as the use of $D^{(0,\xi)}_{\mu\nu}(p)$ for the
QED$_{2+1}$ propagator does not lead to any infrared divergency one can
expect the independence of the physical sector on $\xi$, also the unitarity
in this sector may be demonstrated directly in the framework of QED$_{2+1}$.
We shall justify this suggestion in the next section.

\section{Gauge invariance in the $\xi$--gauge}

Let us consider the Lagrangian

        \be
        \cL_{(0,\xi)} = -\frac14 F_{\mu\nu}F^{\mu\nu} +
        \frac{\theta}4 \ve_{\mu\nu\alpha}F^{\mu\nu}A^\alpha
        - \frac1{2\xi} (\d_\mu A^\mu)^2+ B\d_\mu A^\mu -
        \frac{\xi}{2\theta^2}
        \d_\mu B\d^\mu B +
        \bar{\psi}
        (\imath \hat{\partial} + e \hat{A}   -M)\psi~~~,
        \label{l_xi}
        \ee

\noi
which we shall refer to as Lagrangian of QED$_{2+1}$ in $\xi$--gauge.
The free propagator of the gauge field $A_\mu$ in this model is the following

        \be
        {D}^{(0,\xi)}_{\mu\nu}(p)=
        -\frac{\i}{p^2-\theta^2}
        \left( g_{\mu\nu} - (1-\xi)\frac{p_\mu p_\nu}{p^2}
                + \i \theta \ve_{\mu\nu\alpha}\frac{p^\alpha}{p^2}
                                                          \right)~~~,
        \label{d_xi}
        \ee

\noi
and at $\xi=0$ it transforms to the propagator of QED$_{2+1}$ in the
transversal gauge. Obviously, the main infrared singularity of the propagator
${D}^{(0,\xi)}_{\mu\nu}(p)$ is independent of $\xi$. To calculate the
spectrum of the gauge field modes we should solve the free equations of motion

        \be
        \left(
        \begin{array}{c}
        A_{\mu}\\
        {}\\
        B
        \end{array}
                                                \right)=
        \sum_{i=1}^4 \int d\frac{\bp}{2\pi\sqrt{2\omega_i}}
        (e^{-\i p^{(i)} x}e^{(i)}_l   a_{i}(\bp) + {\rm h.c.})~~~,
        \label{polar_3}
        \ee

        \be
        p_\mu^{(i)}=(\omega_i,\bp)~~~,~~~
        \omega_{1,2}=\sqrt{\bp^2+\theta^2}~~~,~~~
        \omega_{3,4}=|\bp|~~~,
        \ee

        \be
        e^{(1)}_l  = \frac\omega{\theta|\bp|}
        \left(
        \begin{array}{c}
        p^{(1)}_\mu - \delta_{\mu 0} \frac{\theta^2}{\omega_1}
        +\i \frac\theta{\omega_1}\varepsilon_{\mu\alpha0}p^{(1)\alpha}
        \\
        {}\\
        {0}
        \end{array}
                                                \right)~~~,~~~
        e^{(2)}_l  = \frac{\sqrt{\xi}}{\theta}
        \left(
        \begin{array}{c}
        {p_\mu^{(2)}}
        \\
        {}\\
        {-\i \frac\theta\xi}
        \end{array}
                                                \right)~~~,
        \ee

\medskip
        \be
        e^{(3)}_l  = \frac{1}{\sqrt{\theta|\bp|}}
        \left(
        \begin{array}{c}
        p^{(3)}_\mu
        \\
        {}\\
        0
        \end{array}
                                                \right)~~~,~~~
        e^{(4)}_l  = - \frac{1}{\sqrt{\theta|\bp|}}
        \left(
        \begin{array}{c}
        \frac{|\bp|(1+\xi)}{2\theta}p^{(4)}_\mu +\i \ve_{\mu\a0}p^{(4)\alpha}
        \\
        {}\\
        \i |\bp|
        \end{array}
                                                \right)~~~,
        \ee

\normalsize

\noi
where the creation and annihilation operators are subject to the following
commutation relations:

        \be
        [a^{(1)}_\bp,a^{(1)\dagger}_\bq]=
        [a^{(2)}_\bp,a^{(2)\dagger}_\bq]=
        [a^{(3)}_\bp,a^{(4)\dagger}_\bq]=
        [a^{(4)}_\bp,a^{(3)\dagger}_\bq]= \de(\bp-\bq)~~~,
        \label{comm2}
        \ee

\noi
other commutators vanish.

As we shall demonstrate below just one mode of four, that of the polarization
 $e^{(1)}_{l}$, is physical. It does not depend on  $\xi$ and coincides
with the physical mode in  $\a$-gauge. One can obtain the Ward identities in
this model. For the  generating functional of Green functions $W$ they are the
following:

        \be
        \left(
        \d_\mu \frac{\de}{\de I^\mu(x)} +
        \frac{\xi}{\theta^2} \Box \frac{\de}{\de I_B(x)}
                                \right)W +I_B(x)=0~~~,
        \label{w1}
        \ee

        \be
        \left(
        -\frac{\Box+\theta^2}{\xi}\d_\mu \frac{\de}{\de I^\mu(x)}
        + \i e \left(
        \eta(x)\frac{\de}{\de \eta(x)}
        - \bar{\eta}(x)\frac{\de}{\de \bar{\eta}(x)}
                                \right)\right)W
        -\frac{\theta^2}{\xi} I_B(x) - \d_\mu I^\mu(x)=0~~~,
        \label{w2}
        \ee

        \bea
        \lefteqn{\frac{\d W}{\d\xi} =\frac12 \int d\!x
        \left[
        \frac1{\xi^2}\left(\d_\mu \frac{\de W}{\de I^\mu(x)} \right)^2
        + \frac1{\theta^2} \frac{\de W}{\de I_B(x)}\Box
        \frac{\de W}{\de I_B(x)}                \right]}\nonumber\\
        &&
        -\frac{e}{2\xi}\int d\!x d\!y {\rm D}_{(0)}(x-y)
        \d_\mu \frac{\de}{\de I^\mu(x)}
        \left(
        \eta(y)\frac{\de}{\de \eta(y)}
        - \bar{\eta}(y)\frac{\de}{\de \bar{\eta}(y)}
                                \right)W
        \label{w3}
        \eea

\noi
($I_B$ is the source to the field $B$).

For the generating functional of vertex functions $\bar{\Gamma}$,

        \be
        \bar{\Gamma} \equiv \Gamma + \frac1{2\xi} (\d A)^2 -
              B  \d_\mu A^\mu  - \frac{\xi}{2\theta^2} B \Box B~~~,
        \ee

\noi
the Ward identities become

        \be
        \frac{\de \bar{\Gamma}}{\de B} = 0~~~,
        \label{w4}
        \ee

        \be
        \d_\mu \frac{\de \bar{\Gamma}}{\de A_\mu(x)}=
        \i e
        \left(
        \psi(x)\frac{\de}{\de \psi(x)}
        - \bar{\psi}(x)\frac{\de}{\de \bar{\psi}(x)}
                                \right)\bar{\Gamma}~~~,
        \label{w5}
        \ee

        \be
        \frac{\d  \bar{\Gamma}}{\d\xi}=
        \frac{e}{2\xi}\int d\!x d\!y {\rm D}_{(0)}(x-y)
        \left(
        \d_\mu \frac{\de^2W}{\de I^\mu(x)\de \eta(y)}
        \frac{\de \bar{\Gamma}}{\de \psi(y)}
        - \d_\mu \frac{\de^2W}{\de I^\mu(x)\de \bar{\eta}(y)}
        \frac{\de \bar{\Gamma}}{\de \bar{\psi}(y)}
                                \right)~~~.
        \label{w6}
        \ee

We shall show below that the generating functional of vertex functions
 $\bar{\Gamma}$ in this model on the mass shell exists for all the modes and
the arguments underlying the equivalency theorem are valid here, too.
Therefore, the transversality of $\bar{\Gamma}$ in photon momenta on
the fermion mass shell (f.m.s) follows from Eqs. (\ref{w1}), (\ref{w2}) or
Eqs. (\ref{w4}), (\ref{w5}),

        \be
        \frac{\de \bar{\Gamma}}{\de B} =
        \left. \d_\mu \frac{\de \bar{\Gamma}}{\de A_\mu(x)}
        \right|_{f.m.s.} = 0~~~,
        \label{w7}
        \ee

\noi
and independence of $\bar{\Gamma}$ on parameter $\xi$ on the fermion mass
shell

        \be
        \left. \frac{\d \bar{\Gamma}}{\d \xi}\right|_{f.m.s.} = 0~~~
        \label{w8}
        \ee

\noi
appears from \Eq{w3} or \Eq{w6}.

It follows from \Eq{w7} that the $S$-matrix depends on two creation and
annihilation operators out of four, namely $a^{(1)},a^{(1)\dagger}$ and
$a^{(4)},a^{(4)\dagger}$. Due to commutation relations \Eq{comm2} this
implies that the $S$-matrix is unitary in the physical subspace comprising
fermion and photon physical mode with the polarization $e^{(1)}_l$.  Then, it
follows from \Eq{w8} that the matrix elements are independent of $\xi$ in
the physical subspace. In particular, the physical mass of the fermion and
propagating photon mode are $\xi$-invariant. Consider one-loop correction to
the fermion mass as an illustration:

        \be
        {\Sigma}_1 (p) =
        {\Sigma}^{(1)}_1 (p) + {\Sigma}^{(\xi)}_1(p)
        \ee

        \be
        {\Sigma}^{(1)}_1 (p)=
        -\i e^2 \int \frac{d\!k}{(2\pi)^3}
        \g^\mu\frac{(\hat{p}-\hat{k}+M)}{((p-k)^2-M^2)(k^2-\theta^2)}
        \g^\nu
        \left(
        g_{\mu\nu} + \i \theta \ve_{\mu\nu\alpha}\dfrac{k^\alpha}{k^2}
                                                          \right)
        \ee

        \bea
        \lefteqn{{\Sigma}^{(\xi)}_1 (p)=
        -\i e^2 (1-\xi)\int \frac{d\!k}{(2\pi)^3}
        \frac{\hk(\hat{p}-\hat{k}+M)\hk}{k^2((p-k)^2-M^2)(k^2-\theta^2)}=}
        \nonumber\\
        &&
        -\i e^2 (1-\xi)(\hp -M)\int \frac{d\!k}{(2\pi)^3}
        \frac{(\hat{p}-\hat{k}+M)\hk}{k^2((p-k)^2-M^2)(k^2-\theta^2)}
        \eea

${\Sigma}^{(\xi)}_1 (p)$  vanishes on the fermion mass shell and the
only contribution to the mass renormalization comes from
${\Sigma}^{(1)}_1 (p)$, which is independent of $\xi$ and coincides
with the mass renormalization in the transversal gauge.

Let us prove the existence of vertex functions and validity of the
equivalency theorem. The most  infrared singular term in the photon
propagator  \Eq{d_xi} at any $\xi$ is

        \be
        \frac{\theta}{p^2-\theta^2}\cdot \frac{\ve_{\mu\nu\lambda}p^\lambda}
        {p^2}~~~,
        \nonumber
        \ee

\noi
i.e. the propagator in the infrared region is effectively
proportional to ${\ve_{\mu\nu\lambda}p^\lambda}/{p^2}$.
Let us suppose that a diagram contains a closed fermion loop as a subdiagram
(or it is just a closed fermion loop). We shall denote such a loop with $n$
external photon lines as $\Pi_{\mu_1\dots\mu_n}(p_1,\dots,p_n)$. Due to the
transversality of the diagram in any photon momenta,

        \be
        p^{\mu_i}\Pi_{\mu_1\dots\mu_i\dots\mu_n}(p_1,\dots,p_n)=0~~~,
        \label{transverse}
        \ee

\noi
$\Pi_{\mu_1\dots\mu_n}(p_1,\dots,p_n)$ is proportional to {\it
every} photon momentum for  $n \ge 3$. Therefore, infrared singularities of
photon propagators outgoing the fermion loop with more than two
external lines are completely suppressed. For $n=2$ (the polarization
diagram) it follows from \Eq{transverse} that

        \be
        \Pi_{\mu\nu}=(p^2 g_{\mu\nu}-p_\mu p_\nu) A
        +\i \ve_{\mu\nu\lambda}p^\lambda B~~~,
        \ee

\noi
where $A$ and $B$ have finite limit at $p \to 0$, thus the effective photon
line resulting after insertion of the polarization diagrams is proportional
to ${\ve_{\mu\nu\lambda}p^\lambda}/{p^2}$ again in the infrared limit.
Therefore, the (sub)diagrams with closed fermion lines are infrared innocuous
and internal photon lines may be considered as effective ones.

Diagrams left after exclusion of fermion loops are non-closed fermion lines
with outgoing and incoming internal and
external photon lines. One may suppose that each loop momentum integration
is carried out over some internal photon momentum. Since the internal photon
line has infrared singularity $1/k$, infrared divergency of the diagram may
take place when the diagram contains two other lines with propagators $\sim
1/k$.  First, suppose the diagram has no external lines. Since the photon
line is tied to two fermion lines, the divergency may take place if only
terms nonvanishing in $k \to 0$ limit in the denominators of the fermion
propagators are cancelled due to the mass shell condition. Thus the infrared
divergency may arise just in the following type of integrals:

\begin{equation}
\int \!dk
\begin{picture}(100.00,40.00)
\unitlength=1.00mm
\linethickness{0.4pt}
\put(-30,-18){
\put(34.00,25.){\line(1,0){30.}}
\put(39.00,25.){\vector(1,0){1.}}
\put(59.00,25.){\vector(1,0){1.}}
\put(34.00,15.){\line(1,0){30.}}
\put(39.00,15.){\vector(1,0){1.}}
\put(59.00,15.){\vector(1,0){1.}}
\put(70.00,20.00){\oval(15.,15.)}
\put(74.00,27.){\line(1,1){7.}}
\put(78.00,20.){\line(1,0){10.}}
\put(74.00,13.){\line(1,-1){7.}}
\put(44,15.75){\oval(1.5,1.5)[r]}
\put(44,17.25){\oval(1.5,1.5)[l]}
\put(44,18.75){\oval(1.5,1.5)[r]}
\put(44,20.25){\oval(1.5,1.5)[l]}
\put(44,21.75){\oval(1.5,1.5)[r]}
\put(44,23.25){\oval(1.5,1.5)[l]}
\put(44,24.75){\oval(1.5,1.5)[br]}
\put(45.,25.){\circle*{1.}}
\put(44.,15.){\circle*{1.}}
\put(37.00,28.){\makebox(0,0)[cc]{{\small $p_1,\a_1$}}}
\put(37.00,12.){\makebox(0,0)[cc]{{\small $p_2,\a_2$}}}
\put(70.00,20.){\makebox(0,0)[cc]{$T$}}
\put(48.00,20.){\makebox(0,0)[cc]{{\small $k$}}}
\put(55.00,28.){\makebox(0,0)[cc]{{\small $p_1+k$}}}
\put(55.00,12.){\makebox(0,0)[cc]{{\small $p_2-k$}}}
}
\end{picture}~~~~~~~~~~~~~~~~~,
\end{equation}

\vspace{1cm}
\noi
$T$ has no infrared singularities in $k$. Potentially divergent part of the
integral is

        \bea
        \lefteqn
        {\int \!dk
        \frac{(\gamma^\mu(\hp_1+M))_{\a_1\beta_1}
        (\gamma^\nu(\hp_2+M))_{\a_2\beta_2}
        \ve_{\mu\nu\l}k^\l}
        {((p_1+k)^2-M^2)((p_2-k)^2-M^2)k^2}
        T_{\beta_1\beta_2}=}\\[2mm]
        &&
        \int \!dk
        \frac{((\hp_1-M)\gamma^\mu- 2p_1^\mu)_{\a_1\beta_1}
        ((\hp_2-M)\gamma^\nu- 2p_2^\nu)_{\a_2\beta_2}
        \ve_{\mu\nu\l}k^\l}
        {(k^2+2p_1k+\de_1)(k^2-2p_2k+\de_2)k^2}
        T_{\beta_1\beta_2}=\nonumber\\ [2mm]
        &&
        \int \frac{\!dk}
        {(k^2+2p_1k+\de_1)(k^2-2p_2k+\de_2)k^2}
        [((\hp_1-M)\gamma^\mu)_{ \a_1\beta_1}
        ((\hp_2-M)\gamma^\nu)_{ \a_2\beta_2} -  \nonumber \\[2mm]
        &&
        ~~~~~~~~~~~~2((\hp_1-M)\gamma^\mu)_{ \a_1\beta_1}
        p_2^\nu\de_{\a_2\beta_2} -
        2p_1^\mu \de_{\a_1\beta_2}((\hp_2-M)\gamma^\nu)_{ \a_2\beta_2}]
        \ve_{\mu\nu\l}k^\l
        T_{\beta_1\beta_2}                      \nonumber
        \eea

        \be
        \de_1=p^2_1-M^2~~~,~~~\de_2=p^2_2-M^2~~~.
        \nonumber
        \ee
Since in the vicinity of the mass shell one has

        \be
        \hp_1 - M= \frac{\de_1}{2M}~~~,~~~\hp_2 - M= \frac{\de_2}{2M}~~~,
        \nonumber
        \ee

\noi
the integral behaves as $\de \ln \de$ and it is not singular on the mass shell.

Then, suppose the diagram contains external photon lines. Potentially
dangerous are the following diagrams:

\begin{equation}
\int \!dk
\begin{picture}(100.00,40.00)
\unitlength=1.00mm
\linethickness{0.4pt}
\put(-30,-18){
\put(34.00,25.){\line(1,0){30.}}
\put(47.00,25.){\vector(1,0){1.}}
\put(39.00,25.){\vector(1,0){1.}}
\put(59.00,25.){\vector(1,0){1.}}
\put(34.00,15.){\line(1,0){30.}}
\put(39.00,15.){\vector(1,0){1.}}
\put(59.00,15.){\vector(1,0){1.}}
\put(70.00,20.00){\oval(15.,15.)}
\put(74.00,27.){\line(1,1){7.}}
\put(78.00,20.){\line(1,0){10.}}
\put(74.00,13.){\line(1,-1){7.}}
\put(44,15.75){\oval(1.5,1.5)[r]}
\put(44,17.25){\oval(1.5,1.5)[l]}
\put(44,18.75){\oval(1.5,1.5)[r]}
\put(44,20.25){\oval(1.5,1.5)[l]}
\put(44,21.75){\oval(1.5,1.5)[r]}
\put(44,23.25){\oval(1.5,1.5)[l]}
\put(44,24.75){\oval(1.5,1.5)[br]}
\put(52,25.75){\oval(1.5,1.5)[r]}
\put(52,27.25){\oval(1.5,1.5)[l]}
\put(52,28.75){\oval(1.5,1.5)[r]}
\put(52,30.25){\oval(1.5,1.5)[l]}
\put(52,31.75){\oval(1.5,1.5)[r]}
\put(52,33.25){\oval(1.5,1.5)[l]}
\put(52,34.75){\oval(1.5,1.5)[br]}
\put(52.,25.){\circle*{1.}}
\put(45.,25.){\circle*{1.}}
\put(44.,15.){\circle*{1.}}
\put(37.00,28.){\makebox(0,0)[cc]{{\tiny $p_1,\a_1$}}}
\put(37.00,12.){\makebox(0,0)[cc]{{\tiny $p_2,\a_2$}}}
\put(70.00,20.){\makebox(0,0)[cc]{{$T$}}}
\put(48.00,33.){\makebox(0,0)[cc]{{\small $p$}}}
\put(48.00,20.){\makebox(0,0)[cc]{{\small $k$}}}
\put(46.50,28.){\makebox(0,0)[cc]{{\tiny $p_1+k$}}}
\put(60.00,28.){\makebox(0,0)[cc]{{\tiny $p_1+k+p$}}}
\put(55.00,12.){\makebox(0,0)[cc]{{\tiny $p_2-k$}}}
}
\end{picture}~~~~~~~~~~~~~~~~~~~~~~~~~~.
\end{equation}

\vspace{1cm}
The denominator of the additional fermion propagator is

        \be
        (p_1+k+p)^2-M^2=k^2+2pk+p^2+\de_1+2p_1p+2kp_1
        \ee

\noi
and even if the  massless   photon  $(p^2=0)$ is radiated it has
infrared regularization $2p_1p$. Therefore, infrared singularities of
diagrams containing photon radiation are the same as those of diagrams without
radiation and all these diagrams are infrared finite.

Thus we have shown that the vertex functions of  QED$_{2+1}$ in $\xi$-gauge
has no singularities at any $\xi$. In particular, it implies that the vertex
functions in the Proca model have finite limit at $m\to0$ (we have used this
in the previous section).

Let us prove \Eq{w7} and \Eq{w8} now. We shall consider just \Eq{w8}, since
\Eq{w7} may be proven along the same lines (this approach is also valid in
proving the similar equalities in the Proca model). It is more convenient
to demonstrate that the last term in \Eq{w3} is vanishing on the mass shell.
That would mean that \Eq{w8} is satisfied (we are following the approach
elaborated in \refc{KT}).

The last term in the right-hand side of \Eq{w3} has
the following structure:

\medskip
\begin{equation}
\int \!dp \psi_{in}(p)\vec{K}(p)\int\! dk
\begin{picture}(100.00,40.00)
\unitlength=1.00mm
\linethickness{0.4pt}
\put(-41,-14){
\put(44.00,15.){\line(1,0){20.}}
\put(59.00,15.){\vector(1,0){1.}}
\put(70.00,20.00){\oval(15.,15.)}
\put(74.00,27.){\line(1,1){7.}}
\put(78.00,20.){\line(1,0){10.}}
\put(74.00,13.){\line(1,-1){7.}}
\put(44.,15.){\circle*{1.}}
\put(70.00,20.){\makebox(0,0)[cc]{{$M$}}}
\put(55.00,12.){\makebox(0,0)[cc]{$p-k$}}
\put(46.,16.5){\circle*{0.5}}
\put(48.,18.2){\circle*{0.5}}
\put(50.,19.3){\circle*{0.5}}
\put(52.,20.4){\vector(2,1){0.5}}
\put(52.,20.4){\circle*{0.5}}
\put(54.,21.4){\circle*{0.5}}
\put(56.,22.3){\circle*{0.5}}
\put(58.,23.1){\circle*{0.5}}
\put(60.,23.9){\circle*{0.5}}
\put(62.,24.5){\circle*{0.5}}
\put(53.00,25.){\makebox(0,0)[cc]{$k$}}
}
\end{picture}~~~~~~~~~~~~~~~~~~~,~~~~~~~~~~~~~~
\label{69}
\end{equation}

\vspace{1cm}
\noi
where the dotted line denotes the propagator

        \be
        \frac{k_\mu}{k^2(k^2-\theta^2)}~~~.
        \label{w10}
        \ee

We have used above the Ward identities for the photon propagator valid both
for free and exact functions:

        \be
        k^\mu {D}_{\mu\nu} = -\i \frac{\xi k_\nu}{k^2-\theta^2}~~~.
        \nonumber
        \ee

\noi
The operator $K(p)$ in \Eq{69} is

        \be
        K(p)= \hp -M~~~,
        \nonumber
        \ee

\noi
$\psi_{in}(p)$ is a solution of the free equation of motion:

        \be
        K(p)\psi_{in}(p)= 0~~~,~~~\psi_{in}(p) \sim \de (p^2- M^2)~~~.
        \nonumber
        \ee

\noi
As $K(p)$ acts rightwards on the function with no
$1/(\hp -M)$ singularities, \Eq{69} vanishes.

Let the diagram \Eq{69} be one-particle irreducible. In this case $K(p)$
operates on a function of the form

        \be
        \int \! dk \frac{(\hp-\hk+M))k_\mu}
        {(k^2-2pk+\de)k^2(k^2-\theta^2)}M_\mu(k)~~~,~~~ \de=p^2-M^2~~~~,
        \ee

\noi
which is obviously nonsingular on the mass shell $\hp =M, \de = 0$. If the
diagram \Eq{69} is a one-particle reducible one, it has the following
structure:

\begin{equation}
\begin{picture}(100.00,40.00)
\unitlength=1.00mm
\linethickness{0.4pt}
\put(-40,-18){
\put(45.00,20.){\line(1,0){17.}}
\put(54.00,20.){\vector(1,0){1.}}
\put(70.00,20.00){\oval(15.,15.)}
\put(74.00,27.){\line(1,1){7.}}
\put(77.00,20.){\line(1,0){10.}}
\put(74.00,13.){\line(1,-1){7.}}
\put(70.00,20.){\makebox(0,0)[cc]{{$M'$}}}
\put(40.00,20.){\makebox(0,0)[cc]{{$\gamma$}}}
\put(53.00,15.){\makebox(0,0)[cc]{$p$}}
\put(45,20){\circle*{1.}}
\put(35,20){\circle*{1.}}
\put(40.00,20.00){\oval(10.,10.)}
}
\end{picture}~~~~~~~~~~~~~~,~~~~~~~~~~~~~~~~~
\nonumber
\end{equation}

\vspace{0.5cm}
\begin{equation}
\gamma =
\begin{picture}(100.00,40.00)
\unitlength=1.00mm
\linethickness{0.4pt}
\put(-41,-14){
\put(44.00,15.){\line(1,0){20.}}
\put(59.00,15.){\vector(1,0){1.}}
\put(70.00,20.00){\oval(15.,15.)}
\put(76.00,15.){\line(1,0){10.}}
\put(44.,15.){\circle*{1.}}
\put(70.00,20.){\makebox(0,0)[cc]{{$\Gamma$}}}
\put(55.00,12.){\makebox(0,0)[cc]{$p-k$}}
\put(46.,16.5){\circle*{0.5}}
\put(48.,18.2){\circle*{0.5}}
\put(50.,19.3){\circle*{0.5}}
\put(52.,20.4){\vector(2,1){0.5}}
\put(52.,20.4){\circle*{0.5}}
\put(54.,21.4){\circle*{0.5}}
\put(56.,22.3){\circle*{0.5}}
\put(58.,23.1){\circle*{0.5}}
\put(60.,23.9){\circle*{0.5}}
\put(62.,24.5){\circle*{0.5}}
\put(53.00,25.){\makebox(0,0)[cc]{$k$}}
}
\end{picture}~~~~~~~~~~~~~~~,~~~~~~~~~~~~~~~~~~~~~~
\nonumber
\end{equation}

\vspace{1.cm}
\noi
where $M'$ is one-particle irreducible diagram, and \Eq{69} is nonvanishing.
However, since $\gamma$ is nonsingular on the mass shell $\hp =M$,
its contribution is exactly cancelled by the fermion wave function
renormalization (one may find the details in  \refc{KT}).
Note that in  $\a$-gauge the propagator (\ref{w10}) in \Eq{69} should be
changed to

        \be
        \frac{k_\mu}{k^4}~~~,
        \nonumber
        \ee

\noi
thus $K(p)$ in \Eq{69} acts on the function singular on the fermion
mass shell $\hp =M$ and \Eq{69} is nonvanishing (even if vertex functions
exist in $\a$-gauge).

Finally, we have shown that QED$_{2+1}$ is well-defined in
the class of $\xi$-gauges and it contains  $\xi$-independent physical sector
where $S$-matrix is unitary.  In $\xi \to 0$ limit one has exactly the same
expression for the generating functional of vertex functions  $\bar{\Gamma}$
and for operators  $A_\mu$, $\psi$ and $\bar{\psi}$ as in QED$_{2+1}$  in the
transversal gauge, i.e. the transversal gauge belongs to $\xi$-gauges class.
At $\xi=1$ the expression for the free photon propagator is the simplest:

        \be
        {D}^{(1)}_{\mu\nu}=
        -\frac{\i}{p^2 -\theta^2}
        \left(
                g_{\mu\nu}  + \i \theta \frac{\ve_{\mu\nu\alpha}p^\alpha}{p^2}
                                                 \right)~~~,
        \ee

\noi
and the latter gauge is the analogue of the Feynman gauge.

\section*{Acknowledgement}
This work was supported in part by grants RFBR 96-01-00482 and INTAS
96-03-08 (I.~T.) and  RFBR 96-02-16210 and 96-02-16117 (V.~Z.). V.~Z. is
grateful to B.~de Wit for his kind hospitality during the author's stay in
Utrecht University under INTAS grant CT93-0023.

\section*{Appendix}

\setcounter{equation}{0}
\renewcommand{\theequation}{A\arabic{equation}}

We shall discuss here the normalization condition of the
polarization vectors in the Proca model. This question requires
special discussion since the antisymmetric structure associated with the
Chern-Simons coefficient spoils the naive normalization condition
$u^{*(i)}_\alpha u^{(j)\alpha} =\delta_{ij}$. However, correct definition of
the normalization coefficients of the polarization modes is crucial for
reduction formulae and the study of the massless limit of the Proca model.

Let us rewrite the free action for the gauge field $V_\mu$ as follows:

        \be
        S = \frac12 \int d^3x V_\mu \Lambda^{\mu\nu}V_\nu~~~,
        \ee

        \be
        \Lambda^{\mu\nu} = g^{\mu\nu}\d^\alpha\d_\alpha -
        (1-\frac1\alpha)\d^\mu\d^\nu - \theta\ve^{\mu\nu\alpha}\d_\alpha
        +m^2 g_{\mu\nu}~~~.
        \ee

\noi
Then, one may introduce the scalar product of two functions $\phi_\mu(x)$ and
$\psi_\mu(x)$ in a standard way (see, e.g. \cite{DW}):

        \be
        (\phi,\psi) = \i \int d \bx~ J^0(x;\phi,\psi)~~~,
        \ee

\noi
where the current $J^\mu(x;\phi,\psi)$ is:

        \be
        J^\mu(x;\phi,\psi) =\phi^\alpha(j^\mu_{\alpha\beta}\psi^\beta)
        - (j^\mu_{\alpha\beta}\phi^\beta)\psi^\alpha~~~,
        \ee

        \be
        j^\mu_{\alpha\beta} = g_{\alpha\beta}\d^\mu - \frac12(1-\frac1\alpha)
        (g_\alpha^\mu\d_\beta + g_\beta^\mu\d_\alpha) -
        \frac\theta2\ve^\mu_{~\alpha\beta}~~~,
        \ee

\noi
and its derivative $\d_\mu J^\mu(x;\phi,\psi)$ may be written as follows

        \be
        \d_\mu J^\mu(x;\phi,\psi) =
        \phi_\mu\Big(\Lambda^{\mu\nu}\psi_\nu\Big)
        -\Big(\Lambda^{\mu\nu}\phi_\nu\Big) \psi_\mu ~~~.
        \ee

\noi
Therefore, $\d_\mu J^\mu(x;\phi,\psi)$ vanishes on solutions of the free
equation of motion. The zeroth component of the current is equal to

        \be
        J^0(x;\phi,\psi)= {P}_{(\phi)}^\mu  \psi_\mu
        - \phi_\mu {P}_{(\psi)}^\mu ~~~,
        \ee

\noi
with ${P}_{(\phi)}^\mu $ being the momentum canonically conjugated to field
$\phi_\mu$ for  action $S$.

The gauge field $V_\mu(x)$ may be decomposed in the polarization vectors

        \be
        V_\mu(x)= \sum_{i=1}^3
        \int d\bp \Big[ e^{(i)}_{p,\mu}(x) a^{(i)}_p + h.c.\Big]~~~,
        \ee

        \be
        e^{(i)}_{p,\mu}(x)= \dfrac1{2\pi\sqrt{2\omega_i}}
        e^{-\i p^{(i)} x}u^{(i)}_{\mu}(\bp)~~~,
        ~~~p_\mu^{(i)} = (\omega_i,\bp)~~~.
        \ee

Vectors $e^{(i)}_{p,\mu}(x)$ are subjects to two conditions.
First, they are solutions of the free equation of motion,

        \be
        \Lambda^{\mu\nu}e^{(i)}_{p,\mu}(x)=0~~~.
        \ee

\noi
(in the Proca model the latter condition defines the polarizations vectors
up to normalization coefficients).

Second, polarization vectors satisfy normalization conditions,

        \be
        (e_p^{(i)},e_q^{(j)})=0~~~,~~~
        (e_p^{(i)*},e_q^{(j)})=\delta_{ij}\delta(\bp-\bq)~~~.
        \ee

In this case

        \be
        a_p^{(i)}=(e_p^{(i)*},V)~~~,~~~
        a_p^{\dagger(i)}=(e_p^{(i)},V)
        \ee

The normalization conditions (A11), (A12) may be derived also by using
canonical commutation relations

        \be
        \left.[{P}_{\mu}(x),V_\nu(y)]\right|_{x_0=y_0} =
        -\i g_{\mu\nu} \delta({\bf x}-{\bf y})~~~,
        \ee

        \be
        [a_{(i)}(\bp),a^\dagger_{(j)}(\bq)] =
        \delta_{ij}\delta(\bp-\bq)~~~,~~~ i=1,2~~~
        [a_{(3)}(\bp),a^\dagger_{(3)}(\bq)] = -\delta(\bp-\bq)~~~.
        \ee

\noi
Other  commutators vanishe,
$P_\mu=\d L/\d {\dot V_\mu}$
are  momenta canonically conjugated $V_\mu$.

Let us emphasize that the thus defined polarization vectors enter the matrix
elements of the scattering  matrix when these are expressed through the Green
functions  (reduction formulae):

        \bea
        \lefteqn{\langle\dots|S|p,i;\dots\rangle = \langle\dots |S
        a^{\dagger(i)}_p|\dots\rangle=}\\ &&
        \i \int d^3x e^{(i)}_{p,\mu}(x)
        \vec{\Lambda}^{\mu\nu} \langle0|TV_\nu(x)\dots|0\rangle= \int d^3x
        e^{(i)}_{p,\mu}(x) \Gamma^{\mu\dots}(x,\dots)\nonumber
        \eea

\end{document}